\documentclass[12pt]{iopart}
\usepackage{iopams,bm,verbatim}
\def\a{\alpha}
\def\b{\beta}

\def\d{\delta}
\def\e{\epsilon}
\def\diver{\textrm{div}}
\def\curl{\textrm{curl}}
\newcommand{\w}[1]{\bm{#1}} 

\begin{document}

\title{Purely radiative perfect fluids with degenerate shear tensor.}
\author{H.R.~Karimian, N.~Van den Bergh and L.~De Groote}

\address{Department of Mathematical Analysis IW16, Ghent University, Galglaan 2, 9000 Ghent, Belgium}

\begin{abstract}
We consider non-rotating geodesic perfect fluid spacetimes which are "purely radiative" in the sense that the gravitational field satisfies the covariant transverse conditions $\diver \mathbf{H}=\diver \mathbf{E}=0$. We show that when the shear tensor $\w{\sigma}$ is degenerate, $\mathbf{H}$, $\mathbf{E}$ and $\w{\sigma}$ necessarily commute and hence the resulting spacetimes are hypersurface homogeneous of Bianchi class $A$ (modulo some purely electric exceptions).
\end{abstract}

\section{Introduction}
Although gravitational waves in perfect fluid spacetimes are usually studied via
transverse traceless tensor perturbations on a Friedman-Lema\^\i tre-Robertson-Walker (FLRW) background, there is renewed interest in the so-called covariant approach~\cite{Ehlers, Hawking, Ellis, Ellisbook2} in which gravitational radiation is described via the electric and magnetic
parts of the Weyl tensor, which represents the locally free part
of the gravitational field. Perfect fluid spacetimes in which the covariant
transverse conditions $\diver \mathbf{E} = \diver \mathbf{H} = 0$ hold at the non-perturbative level have been called \emph{purely radiative}~\cite{Sopuerta} and were shown to be very restricted in the sense that they would have to obey two non-terminating chains of integrability conditions (see for example \cite{vanElst, Mars}). All the evidence so far seems to indicate~\cite{Maartens2} that these conditions only hold in models with special symmetries, such as the spatially homogeneous spacetimes of Bianchi class A~\cite{Ellis3, Wainwright}. In models with realistic inhomogeneity, the gravito-magnetic field would therefore necessarily be non-transverse at second and higher order.
\par Recently we discussed purely radiative irrotational dust spacetimes for which the magnetic part of Weyl tensor is diagonal in the shear-electric eigenframe~\cite{dust}, as well as their generalization to purely radiative and geodesic perfect fluids with non-constant pressure~\cite{PurelyPF}. In the latter paper, hereafter referred to as paper I, we showed that the Bianchi class A perfect fluids with non-constant pressure could be uniquely characterized (modulo some purely electric exceptions) as those purely radiative models for which the magnetic and electric part of the Weyl tensor and the shear tensor are simultaneously diagonalizable. For the case of constant pressure the same conclusion holds provided the spatial gradient of the matter density $\rho$ vanishes: the velocity then becomes orthogonal to the $\rho =$ constant hypersurfaces, and hence the flow is irrotational~\cite{Synge, RaychMaity}.
Our aim was to show that this analysis could be generalized to the case where $\mathbf{H}$, $\mathbf{E}$ and $\w{\sigma}$ are not necessarily simultaneously diagonalizable. The initial steps of this programme looked hopeful, in the sense that rational expressions could be constructed for all relevant quantities appearing in the description of the problem. However the sheer size of these expressions made any subsequent investigation of the ensuing integrability conditions virtually impossible. For technical reasons we therefore assume in the present contribution that the shear tensor is degenerate. A similar, but altogether different, assumption was made in \cite{Mustapha}, where it was assumed that all kinematical variables, together with the electric and magnetic parts of the Weyl curvature were rotationally symmetric about a common spatial axis. This assumption gives rise to the so called partially locally rotationally symmetric (PLRS) cosmologies. The intersection of the PLRS cosmologies with our present class of models (i.e.~those cosmologies having $\diver \mathbf{H}=\diver \mathbf{E}=0$ and a degenerate shear tensor), consists precisely of the LRS Bianchi cosmologies and the Kantowski-Sachs perfect fluids  discussed in paper I.

\par In \S 2 we give an overview of the results of paper I which remain applicable to the case under study, while in \S 3 and \S 4 we present a proof of the main result.

\section{Main equations}
 We will follow the notations and conventions of paper I. All calculations were performed using the OFRAME package~\cite{Vanden} written in the symbolic computation language MAPLE and which is available from the authors.

The energy momentum tensor of a perfect fluid has the form:
\begin{equation}
    T_{ab} = (\rho + p)u_au_b+ pg_{ab}, \quad u_au^a=-1
\end{equation}
where $\rho$ is the energy density in the rest-frame, $p$ the pressure of the fluid and $\mathbf{u}$ the normalized $4$-velocity. We also define  $\mathbf{E}$ and $\mathbf{H}$ as the electric and magnetic part of the Weyl tensor with respect to the congruence $u^a$:
\begin{equation}\label{EenH}
E_{ab}\equiv C_{acbd}u^c u^d   \quad  H_{ab}\equiv \frac{1}{2}{\varepsilon_{acd}}{C^{cd}}_{be}u^e,
\end{equation}
where $\varepsilon_{abc}=\eta_{abcd}u^d$ is the spatial projection
of the spacetime permutation tensor $\eta_{abcd}$.
\par From the `$\diver \mathbf{H}$' Bianchi identity (see the appendix of~\cite{PurelyPF}) and the assumption $\diver \mathbf{H}=0$ we obtain $[\w{\sigma}, \mathbf{E}] =0$, implying the existence of a common $\w{\sigma}$ and $\mathbf{E}$ eigenframe, which will be used henceforth. From the degeneracy of the shear tensor one can easily infer then that $\mathbf{E}$ must be degenerate too~\cite{PurelyPF}. Therefore we may assume that:
\begin{equation}\label{degen}\left.\begin{array}{lll}
\w{\sigma} &=&\textrm{diag}(-2\sigma,\sigma,\sigma)\\
\mathbf{E} &=&\textrm{diag}(-2E,E,E),\\
\end{array}\right.
\end{equation}
where $\sigma$ and $E$ are non-zero, as otherwise the space-time is FLRW.
As explained in~\cite{PurelyPF}, because all off-diagonal elements of $\mathbf{E}$ vanish, we have $\Omega_2=\Omega_3=0$. By choosing the tetrad to be Fermi propagated we may also assume $\Omega_1=0$. Herewith the rotation rate of the tetrad about the $\mathbf{e}_1$-axis is fixed up to rotations
\[\begin{array}{lll}
\mathbf{e}_2 &\rightarrow& \mathbf{e}_2 \cos \alpha + \mathbf{e}_3 \sin \alpha \\
\mathbf{e}_3 &\rightarrow& -\mathbf{e}_2 \sin \alpha +\mathbf{e}_3 \cos \alpha ,
\end{array}
\]
satisfying $\partial_0 \alpha=0$.
With $\dot{\w{u}}=\w{\omega}=0$ the equations (4)-(12) of~\cite{PurelyPF} simplify to the following ($\mathbf{Z}$ is the spatial gradient of the expansion scalar, while $n_{\alpha}, r_{\alpha}, q_{\alpha}$ are defined in terms of the rotation coefficients $\gamma^\a_{\b\d}=\epsilon_{\b\d\e}n^{\e\a}+\delta^\a_\d a_\b -
\delta^\a_\b a_\d$ by $n_{\alpha +1\ \alpha-1}=(r_\alpha+q_\alpha)/2$, $a_\alpha=(r_\alpha-q_\alpha)/2$ and $n_{\alpha
\alpha}=n_{\alpha+1}+n_{\alpha-1}$):

\begin{equation}\label{dotThetaDotRho}  \begin{array}{lll}
\partial_0\theta&=&-\frac{1}{3}\theta^2-\sigma^2-\frac{1}{2}(\rho+3p)\\
\partial_0\rho&=& -(\rho+p)\theta\\
\partial_0\sigma&=&\sigma (\sigma-\frac{2}{3}\theta)-E\\
\end{array}
\end{equation}
\begin{equation}\label{sigmaD}\left.\begin{array}{lll}\\
\partial_1\sigma&=&-\frac{1}{3}Z_1+\frac{3}{2}\sigma(r_1-q_1)\\
\partial_2\sigma&=&\frac{2}{3}Z_2-3\sigma q_2\\
\partial_3\sigma&=&\frac{2}{3}Z_3+3\sigma r_3\\
\end{array}\right.
\end{equation}
\begin{equation}\label{divE}\left.\begin{array}{lll}\\
\partial_1 E &=& \frac{3}{2}(r_1-q_1)E\\
\partial_2 E &=& 3q_2 E \\
\partial_3 E &=& -3r_3 E\\
\end{array}\right.
\end{equation}
\begin{equation}\label{divrho9}\left.\begin{array}{lll}\\
\partial_1\rho&=&0\\
\partial_2\rho&=&-9(Z_2-3\sigma q_2)\sigma\\
\partial_3\rho&=& -9(Z_3+3\sigma r_3)\sigma\\
\end{array}\right.
\end{equation}
\begin{equation}\left.\begin{array}{lll}\\
H_{11} &=& 3(n_3+n_2)\sigma\\
H_{22} &=& -3n_3\sigma \\
H_{33} &=& -3n_2\sigma\\
\end{array}\right.
\end{equation}
\begin{equation}\label{Hab}\left.\begin{array}{lll}\\
H_{12}&=&-(Z_3+3\sigma r_3)\\
H_{23}&=&-\frac{3}{2}\sigma (r_1+q_1)\\
H_{13}&=&Z_2-3\sigma q_2\\
\end{array}\right.
\end{equation}

while the commutators simplify to:
\begin{equation}\label{commu}\left.\begin{array}{lll}\\
\ [\partial_0,\partial_1] &\equiv& (2\sigma-\frac{1}{3}\theta)\partial_1 \\
\ [\partial_0,\partial_2] &\equiv& -\frac{1}{3}(3\sigma+\theta)\partial_2 \\
\ [\partial_0,\partial_3] &\equiv& -\frac{1}{3}(3\sigma+\theta)\partial_3 \\
\ [\partial_2,\partial_3] &\equiv& (n_2+n_3)\partial_1+q_3\partial_2+r_2\partial_3\\
\ [\partial_3,\partial_1] &\equiv& r_3\partial_1+(n_1+n_3)\partial_2+q_1\partial_3\\
\ [\partial_1,\partial_2] &\equiv& q_2\partial_1+r_1\partial_2+(n_1+n_2)\partial_3.\\
\end{array}\right.
\end{equation}
Acting with $[\partial_0 , \partial_1]$ on $\theta$ and $\rho$ implies $q_1=r_1$ and $Z_1=0$, while
evaluation of the following combination of commutators, $9\sigma[\partial_2,\partial_3]\sigma+[\partial_2,\partial_3]\rho+3\sigma[\partial_2,\partial_3]\theta ,$ leads to
\begin{equation}\label{Z2Z3}
q_2Z_3+r_3Z_2 = 0.
\end{equation}
Furthermore, acting with $[\partial_0,\partial_\alpha]$ ($\alpha=2,3$) on $\theta$ gives us the time evolution of $Z_2$ and $Z_3$:
\begin{equation}\label{Z230}\left.\begin{array}{lll}
\partial_0Z_2=-\frac{9}{2}\sigma Z_2+\frac{45}{2}q_2\sigma^2-\theta Z_2\\
\partial_0Z_3=-\frac{9}{2}\sigma Z_3-\frac{45}{2}r_3\sigma^2-\theta Z_3 ,\\
\end{array}\right.
\end{equation}
which guarantees (cf.~above) that we can rotate the ($e_2,e_3$) basis vectors such that for example $Z_2 = Z_3$ (as $\partial_0 (Z_2/Z_3)=0$). Note that $Z_2$ and $Z_3$ cannot both be zero, as then equation (\ref{Z230}) would imply $q_2 = r_3 =0$, such that by (\ref{divrho9}) the spatial gradient of $\rho$ would vanish. The tetrad could then be further specified by rotating such that $n_{23}=\frac{1}{2}(r_1+q_1)=0$, which by (\ref{Hab}) would take us back to the case $[\w{\sigma} , \mathbf{E}]=[\w{\sigma} , \mathbf{H}]=[\mathbf{E}, \mathbf{H} ] = 0 $~\cite{PurelyPF}. 
So henceforth we work in a fixed frame with $Z_2 = Z_3 \neq 0$ which by (\ref{Z230}) implies $r_3 = -q_2$. Together with (\ref{sigmaD}) the Jacobi equations imply then
\begin{eqnarray}
\partial_0 r_1 &=& (2 \sigma -\frac{1}{3} \theta) r_1 , \\
\label{eq_q2} \partial_0 q_2 &=& -Z_2 +(5\sigma -\frac{1}{3} \theta) q_2.
\end{eqnarray}
Applying now the commutators $[\partial_1,\partial_2]$ and $[\partial_3,\partial_1]$ to $\theta$ and combining the resulting equations, we obtain:
\begin{equation}\label{n1}
 2r_1+2n_1+n_2+n_3=0.
\end{equation}
Propagating $Z_2=Z_3$ along $\w{e}_2$ and $\w{e}_3$ results in
\begin{equation}\label{r2}
Z_2(r_2+q_3)+3\sigma (n_3^2-n_2^2)=0\\
\end{equation}
which together with (\ref{n1}) and the `dot$\mathbf{H}$' equations
\[\dot H_{<ab>}= \curl E_{ab}-\theta H_{ab}+3\sigma_{c<a}{H_{b>}}^c\]
leads to the key equation:
\begin{equation}\label{main}\left.\begin{array}{lll}
(n_2^2-n_3^2)\chi=0\\
\end{array}\right.
\end{equation}
where $\chi\equiv Z_2(2E-9\sigma^2)+45q_2\sigma^3$.
If we propagate $\chi=0$ twice along the fluid flow lines, we re-obtain $Z_2=Z_3=0$. From (\ref{main}) we conclude therefore that $n_2^2-n_3^2=0$ and hence, by (\ref{r2}), $q_3=-r_2$. The resulting cases $n_2+n_3=0$ and $n_2+n_3\neq 0$ (and hence $n_2=n_3 \neq 0$) will be investigated in the following paragraphs.

\section{The case $\mathbf{n_2+n_3=0}$}

The `dot$\mathbf{E}$' equations
\[\dot E_{<ab>}=\curl H_{ab}-\theta E_{ab}+3\sigma_{c<a}{E_{b>}}^c-\frac{1}{2}(\rho+p)\sigma_{ab}\]
imply that also $\mathbf{\curl H}$ is diagonal in the shear eigenframe. Using the off-diagonal components of $\mathbf{\curl H}$ we can propagate $n_2+n_3=0$ along $\w{e}_2$ and $\w{e}_3$ respectively, which leads to
\begin{equation}\label{Z2cond}
(Z_2-4q_2\sigma)(n_2-r_1)=0=(Z_2-4q_2\sigma)(n_2+r_1).
\end{equation}
First notice that $Z_2-4q_2\sigma=0$ would lead to a vanishing spatial gradient of the matter density. In fact, propagating $Z_2-4q_2\sigma=0$ along the fluid flow lines gives $q_2(8 E-7\sigma^2)=0$, after which the propagation of $8 E-7 \sigma^2=0$ along $\w{e}_2$ would yield $q_2\sigma^2=0$ and hence $Z_2=Z_3=0$. From (\ref{Z2cond}) we conclude therefore that $n_2=n_3=r_1=0$.

Propagating $n_2=0$ along $\w{e}_1$ one obtains then
\begin{equation}\label{Eamount}
\ E-\frac{\rho}{3}-2\sigma^2-\frac{\sigma\theta}{3}+\frac{\theta^2}{9}+2r_2q_2=0.\\
\end{equation}
Acting with the commutator $[\partial_0 , \partial_2]$ on $Z_2$ and using (\ref {Eamount}) allows us to eliminate $E$, which implies
\begin{equation}\label{temp2}
-12r_2q_2^3+(2\rho+\frac{141}{2}\sigma^2+2\theta\sigma-\frac{2}{3}\theta^2)q_2^2-\frac{69\sigma Z_2}{2}q_2+5Z_2^2=0\\
\end{equation}
and which also shows that $q_2\neq 0$.
Propagating the latter relation along $\w{e}_3$ and using (\ref{temp2}) to eliminate $\rho$ one finds:
\[\label{vglke}
 (Z_2-3q_2\sigma)(4Z_2-21q_2\sigma)=0.
\]

When we propagate the first factor along the fluid flow lines, we obtain an expression for $\rho$ which, after substitution in (\ref{Eamount}), would lead to the ``anti-Newtonian" result $E=0$, which is impossible \cite{Wylleman}. On the other hand, if we apply the same procedure to the second factor of (\ref{vglke}), we get $\sigma(4Z_2-9q_2\sigma)=0$ which gives a contradiction with $4Z_2-21q_2\sigma=0$.

\section{The case $\mathbf{n_2+n_3\neq 0}$}
We now have $n_3 = n_2\neq 0$, which, when propagated along $\w{e}_1$, implies
\begin{equation}
E=2q_3q_2-2r_1n_2+\frac{1}{3}\rho+2\sigma^2+\frac{1}{3}\sigma\theta-\frac{1}{9}\theta^2-n_2^2-r_1^2.\\
\end{equation}
Acting with the commutator $[\partial_0 , \partial_2]$ on $\rho$, $Z_2$ and $r_1$ we arrive at the following expressions involving the matter density and the pressure:

\begin{eqnarray}\label{vg12}
\nonumber
-\frac{69}{2}q_2\sigma Z_2+15n_2r_1\sigma^2+4n_2r_1\rho-\frac{4}{3}n_2r_1\theta^2+5Z_2^2+\frac{141}{2}q_2^2\sigma^2\\
\nonumber
+12q_2^3q_3-\frac{2}{3}\theta^2q_2^2-12n_2r_1q_2^2+2\rho q_2^2-6n_2^2q_2^2-6r_1^2q_2^2+4n_2r_1\sigma\theta\\
\nonumber
+24n_2r_1q_3q_2+2q_2^2\sigma\theta-12r_1n_2^3-24n_2^2r_1^2-12n_2r_1^3+45\sigma^2r_1^2=0\\
\end{eqnarray}
\begin{eqnarray}\label{vg13}
\nonumber
30Z_2n_2\rho-18r_1Z_2\rho+54r_1^3Z_2-90Z_2n_2^3-18r_1Z_2\sigma\theta-108r_1Z_2q_3q_2\\
\nonumber
+30Z_2n_2\sigma\theta+180Z_2n_2q_3q_2+2025\sigma^3q_2n_2+1377r_1q_2\sigma^3-405r_1Z_2\sigma^2\\
\nonumber
+6r_1Z_2\theta^2-126r_1Z_2n_2^2+18r_1^2Z_2n_2-333Z_2n_2\sigma^2-10Z_2n_2\theta^2=0\\
\end{eqnarray}
and
\begin{eqnarray}\label{pressure}
\nonumber
p=-\frac{27}{2}\sigma^2-6\sigma\theta-4\rho-18q_3q_2+\theta^2+9n_2^2+18r_1n_2+9r_1^2\\
\nonumber
+\frac{189q_2\sigma^3+54q_2\sigma^2\theta+36q_2\sigma\rho+216q_2^2\sigma q_3-12q_2\sigma\theta^2}{2Z_2}\\
-\frac{108q_2\sigma n_2^2+216q_2\sigma r_1n_2+108q_2\sigma r_1^2}{2Z_2}.
\end{eqnarray}
Eliminating $q_3$ from (\ref{vg12}) and (\ref{vg13}) gives the relation
\begin{eqnarray}\label{vg11}
\nonumber
-50Z_2^3n_2+918r_1q_2^3\sigma^3+30Z_2^3r_1+1350\sigma^3q_2^3n_2+1836r_1^2q_2\sigma^3n_2\\
\nonumber
+2700\sigma^3q_2n_2^2r_1+153r_1Z_2\sigma^2q_2^2-900r_1^2Z_2\sigma^2n_2-927Z_2n_2\sigma^2q_2^2\\
\nonumber
-207q_2\sigma Z_2^2r_1+345q_2\sigma Z_2^2n_2+270\sigma^2r_1^3Z_2-594Z_2n_2^2\sigma^2r_1=0.\\
\end{eqnarray}

Notice that $r_1$ has to be different from zero (see  the appendix), while (\ref{eq_q2}) guarantees that $q_2\neq 0$. This enables us to introduce new dimensionless variables,\[ x=\frac{Z_2}{q_2 \sigma} , y=\frac{q_2}{r_1}, h=\frac{\theta}{\sigma}, n=\frac{n_2}{r_1}, r=\frac{r_1}{\sigma}, q=\frac{q_3}{\sigma} \textrm{ and } m=\frac{\rho}{\sigma^2},\] which satisfy the following evolution equations
\begin{equation}\label{e0x}\begin{array}{lll}
\sigma^{-1} \partial_0 x &=& x^2 -[\left( n+1 \right)^2 r^2 - 2\, qyr +(2\,h^2-6\, h-6\,m+153)/18 ]\, x + 45/2\\
\sigma^{-1} \partial_0 n &=& -6\,n\\
\sigma^{-1} \partial_0 r &=& -\left( n+1 \right)^2 {r}^{3}+2\,qy{r}^{2}+ -({h}^{2}-6\,h-3\,m -27)/9\, r
\\
\sigma^{-1} \partial_0 y &=& - \left( x-3 \right)\, y
\end{array}
\end{equation}
and
\begin{equation}\label{e2x}\begin{array}{lll}
\sigma^{-1} \partial_2 x &=& 2\,r [ -\frac{1}{3}\,yx^2 + \left( n+1+2\,y^2 \right)\, xy^{-1} -6\,ny^{-1} - 3\,y ]
\\
\sigma^{-1} \partial_2 n &=& ry\, [ \left(  \frac{1}{2}-\frac{1}{3} \, n - \frac{5}{6}\, n^2  \right)\, x + 4\, n^2 - 2 ] +2\,qn
\\
\sigma^{-1} \partial_2 r &=& \left( 5\,-4\,n-\frac{7}{6}\,x+\frac{5}{6}\,xn \right) {r}^{2}y-2\,qr\\
\sigma^{-1} \partial_2 y &=& \left( \frac{1}{2}\,x-\frac{5}{6}\,xn+4\,n-3\, \right) {y}^{2}r+3\,qy-2\,rn-2\,r.\\
\end{array}\\
\end{equation}
Hereby (\ref{pressure}) was used in order to substitute for all occurrences of the pressure in the right hand sides.
In what follows we will show that the propagation of equations (\ref{vg13}) and (\ref{vg11}) implies that $x$ is constant. Then however (\ref{e2x})(a) and $\partial_0\partial_2x=\partial_0^2\partial_2x=0$ simplify to $(x-6)(x-3)^2y^2+9x=0$ and $\sigma(x-6)(x-3)^3 y^2=0$. This implies $x=0$, which brings us back to the degenerate shear case of paper $I$ and thereby ends the proof of our theorem.

First we rewrite equations (\ref{vg13}) and (\ref{vg11}) as follows
\begin{eqnarray}\label{vg131}
\nonumber
-18\, mx+30\, mxn-18\, hx-108\, qyrx+30\, hxn+180\, xyrnq\\
\nonumber
+54\,{r}^{2}x-90\,x
{r}^{2}{n}^{3}+2025\,n+1377-405\,x+6\,{h}^{2}x-126\,{r}^{2}{n}^{2}x\\
\nonumber
+18\,{r}^{2}nx-333\,nx-10\,xn{h}^{2}
=0\\
\end{eqnarray}
\begin{eqnarray}\label{vg111}
\nonumber
1350\,{y}^{2}n+918\,{y}^{2}-50\,{x}^{3}{y}^{2}n+30\,{x}^{3}{y}^{2}+
1836\,n+2700\,{n}^{2}\\
\nonumber
+153\,{y}^{2}x-927\,x{y}^{2}n-207\,{y}^{2}{x}^{2}
+345\,{y}^{2}{x}^{2}n+270\,x\\
-900\,nx-594\,x{n}^{2}=0.
\end{eqnarray}
Propagation along $\w{e}_2$ of (\ref{vg131}) and (\ref{vg111}) results in a linear homogeneous system in $q$ and $r$, the coefficients of which are polynomials in $n,x$ and $y$. Eliminating $n$ from the determinant of this system and from (\ref{vg111}) yields a polynomial relation in $x$ and $y$:

\begin{eqnarray}\label{eq2na}
\nonumber
P_1(x,y) \equiv  -545500000\,{x}^{14}{y}^{14}+22883025000\,{x}^{13}{y}^{14}
-12001000000\,{x}^{14}{y}^{12} \nonumber \\
+\ldots (76\textrm{ terms}) \ldots +7746297620889600\,{x}^{2}+140466196858798080\,{y}^{2}\nonumber \\
-20656793655705600\,x=0 \nonumber \\
\end{eqnarray}

We construct a second polynomial relation between $x$ and $y$ by propagating equation (\ref{vg111}) along the fluid flow, which yields
\begin{eqnarray}\label{vgll10}
\nonumber
-300\,n{x}^{4}{y}^{2}+250\,{n}^{2}{x}^{4}{y}^{2}+90\,{x}^{4}{y}^{2}-3600\,{n}^{2}{x}^{3}{y}^{2}
+8460\,n{x}^{3}{y}^{2}\\
\nonumber
-3780\,{x}^{3}{y}^{2}+11925\,{n}^{2}{x}^{2}{y}^{2}
-54900\,n{x}^{2}{y}^{2}+28647\,{x}^{2}{y}^{2}\\
\nonumber
+144018\,nx{y}^{2}-69714\,x{y}^{2}
-8748\,{n}^{2}x{y}^{2}+2718\,{n}^{2}{x}^{2}\\
\nonumber
+2970\,{n}^{3}{x}^{2}+810\,{x}^{2}-4050\,n{x}^{2}
+38556\,{y}^{2}
-161028\,n{y}^{2}\\
\nonumber
-12960\,x-49896\,{n}^{3}x+1296\,{n}^{2}x+65880\,nx
+162000\,{n}^{3}\\
\nonumber
-127656\,{n}^{2}-162648\,n+38880=0\\
\end{eqnarray}
Eliminating $n$ from (\ref{vg111}) and (\ref{vgll10}) results then in
\begin{eqnarray}\label{eq2nb}
 \nonumber
P_2(x,y) \equiv 10721500\,{x}^{10}{y}^{6}-380191500\,{x}^{9}{y}^{6}+6061903695\,{x}^{8}{y}^{6}\\
\nonumber
+\ldots (27\textrm{ terms}) \ldots -1402072053120\,x -1800380390400
=0.\\
\end{eqnarray}
 One now can calculate the resultant of  (\ref{eq2na}) and (\ref{eq2nb}) with respect to $y$: this yields a polynomial in $x$ which is not identically 0 and thereby shows that $x$ is a constant.

\section{Conclusion}
A geodesic and non-rotating perfect fluid (where `non-rotating' can be dropped when we have non-constant pressure), with degenerate shear tensor $\w{\sigma}$ and with $\diver \mathbf{E} =\diver \mathbf{H}=0$ has commuting $\mathbf{H}$, $\mathbf{E}$ and $\w{\sigma}$.
By paper I this implies that the resulting spacetimes are spatially homogeneous of Bianchi class A. More particularly, because of the degeneracy of $\w{\sigma}$, they are of type VI$_0$ (in the non-LRS case), or, in the LRS case, of Bianchi types I (VII$_0$), II, VIII or IX. The only exceptions arise in the purely electric case, where also the pseudo-spherically symmetric Bianchi class B type III spacetimes and the Kantowski-Sachs spacetimes are allowed.

\section{Appendix}
\subsection*{the case $r_1=0$}
 Propagation of $r_1=0$ along $\w{e}_2$ gives $n_2(24q_2\sigma-5Z_2)=0$. As the evolution of $24q_2\sigma-5Z_2=0$ along the fluid flow would lead to $\sigma=0$ we have $n_2=0$.
Propagating (\ref{vg12}) along $\w{e}_2$ and substituting $n_2=0$ gives $(Z_2-3q_2\sigma)(4Z_2-21q_2\sigma)=0$.
The first factor must be nonzero, as otherwise (with $r_1=0$) we would have a diagonal $H$. When the second factor is 0 we obtain from (\ref{vg12})
\begin{equation}\label{n2=0}
\frac{435}{16}\sigma^2+12q_3q_2-\frac{2}{3}\theta^2+2\rho+2\sigma\theta=0.\\
\end{equation}
Eliminating $\rho$ from the time evolution of $4Z_2-21q_2\sigma=0$ and (\ref{n2=0}) gives $\sigma=0$.

\section*{References}


\providecommand{\newblock}{}

\end{document}